\documentclass[prl,showkeys,twocolumn,a4paper,amsmath,floatfix]{revtex4}

\usepackage{graphicx}
\usepackage{bm}
\usepackage{array}
\usepackage{booktabs}

\begin{document}


\bibliographystyle{prsty}

\title{Vacancy-stabilized crystalline order in hard cubes}

\author{Frank Smallenburg$^1$}
\author{Laura Filion$^{1,2}$}
\author{Matthieu Marechal$^3$}
\author{Marjolein Dijkstra$^1$}
\affiliation{$^1$ Soft Condensed Matter, Debye Institute for NanoMaterials Science, Utrecht University, 3584 CC Utrecht, the Netherlands\\
$^2$ University Chemical Laboratories, Lensfield Road, Cambridge CB2 1EW, United Kingdom\\
$^3$ Institut f\"ur Theoretische Physik II,
Heinrich-Heine Universit\"at D\"usseldorf,
Universit\"atsstra\ss e 1,
40225 D\"usseldorf,
Germany
}

\date{\today}
\keywords{vacancies | hard polyhedra  | colloids}



\begin{abstract} 
We examine the effect of vacancies on the phase behavior and structure of systems consisting of hard cubes using event-driven molecular dynamics and Monte Carlo simulations. We find a first-order phase transition between a fluid and a simple cubic crystal phase that is stabilized by a surprisingly large number of vacancies, reaching a net vacancy concentration of  $\sim6.4\%$ near bulk coexistence. Remarkably, we find that vacancies increase the positional order in the system.
Finally, we show that the vacancies are delocalized and therefore hard to detect.
\end{abstract}

\maketitle






\label{sec:intro}

The free energy of crystal phases is generally minimized by a finite fraction of point defects like vacancies and interstitials. 
However,  the equilibrium number of such defects in most, if not all, colloidal and atomic/molecular crystals with a single constituent is extremely low.
Exemplarily, for the face-centered cubic crystal of hard spheres, one of the few colloidal
systems where the vacancy and interstitial fractions have been calculated, the fraction 
of vacancies and interstitials is on the order of $10^{-4}$ and $10^{-8}$ respectively, 
near coexistence \cite{defects}.  As such, neither the vacancies nor the interstitials 
strongly affect the phase behaviour and so most studies of crystals 
ignore the effect of these defects.
Nevertheless, vacancies and interstitials have a significant effect on the dynamics in an otherwise perfect crystal,
as the main mechanism for particle diffusion is hopping of particles from filled to empty sites or between interstitial sites.

In this paper, we examine a system of hard cubes where, as we will demonstrate,  one cannot ignore the presence of vacancies. 
Arguably, a cube is one of the simplest non-spherical shapes and the archetype of a space-filling polyhedron.  
Surprisingly, despite the simplicity of this system, we find that the stable ordered phase is strongly affected by the presence of vacancies, 
so much that vacancies actually {\it increase} the positional order and change the 
melting point.  Remarkably, the fraction of vacancies in this system is more than two orders of 
magnitude higher than that of hard spheres or any other known typical, experimentally realizable, single-component atomic or colloidal system, reaching 6.4\% near coexistence.  
Additionally, while purely hard (not rounded) colloidal cubes are yet to be realized, colloidal cubes with various interactions are now a reality
\cite{microcubesynthesis1,nanocubesynthesis1,nanocubesynthesis2,nanocubesynthesis3,rossi2010cubic,nanocubes2011}
and it is likely that hard cubes will be realized in the future. 


Here, we use Monte Carlo (MC) and event driven molecular dynamics (EDMD) 
\cite{edmdanisotropic} simulations 
to examine in detail the effect of vacancies on the equilibrium phase behavior 
of hard cubes. 
The model we study consists of $N$ perfectly sharp hard cubes 
with edge length $\sigma$ in a volume $V$. 
Aside from hard-core interactions which prevent configurations of overlapping cubes, 
the particles do not interact.  
In both types of simulation (MC and EDMD), overlaps are detected using an algorithm based on the 
separating axis theorem (e.g. Ref. \cite{separatingaxis}). More information on the EDMD implementation for cubes
is given in the Supporting Information.

\section{Results}
\subsection{Spontaneous vacancy formation}
The equation of state for a {\it vacancy-free} system of hard cubes has been the subject of a number 
of studies \cite{escobedopolyhedral,escobedoparallelepipeds,jagla_1998} 
and was most recently examined by Agarwal and Escobedo \cite{escobedopolyhedral}. It clearly shows a single, 
first-order phase transition between a fluid and an ordered phase.
The authors of Ref. \cite{escobedopolyhedral} identified the ordered phase at coexistence 
to be a liquid crystal mesophase, i.e. a cubatic phase, which is characterized  by the presence of long-range orientational 
order along three perpendicular axes, but a lack of long-range positional order.
However, the authors noted that finite-size effects made it difficult to
determine the extent of the positional order in their system, and based this
identification on their observation of finite diffusion in the ordered phase.
At high densities, both Refs. \cite{escobedopolyhedral} and \cite{jagla_1998} 
agree that the ordered phase is a simple cubic crystal.





There is no fundamental reason why diffusion cannot occur in a crystal,
hence this is an insufficient criterion for distinguishing between a cubatic
phase and a crystal. 
To study more closely the range of the positional order
along the ordered branch we re-examined the intermediate density region (near
coexistence) using highly efficient EDMD
\cite{edmdanisotropic} simulations allowing us to access system sizes more than an order of
magnitude larger than the ones considered in Ref.
\cite{escobedopolyhedral}. We simulated systems 
of $N=40^3=64000$ particles starting from a  
simple cubic crystal lattice. Coexistence 
between the fluid and ordered phase is observed directly for overall packing fractions $0.455 \leq \eta = N \sigma^3/V \leq 0.480$.
Snapshots of typical configurations are 
shown in the supporting information.  

\begin{figure}[t!]
\begin{center}
\includegraphics[width= 0.3\textwidth]{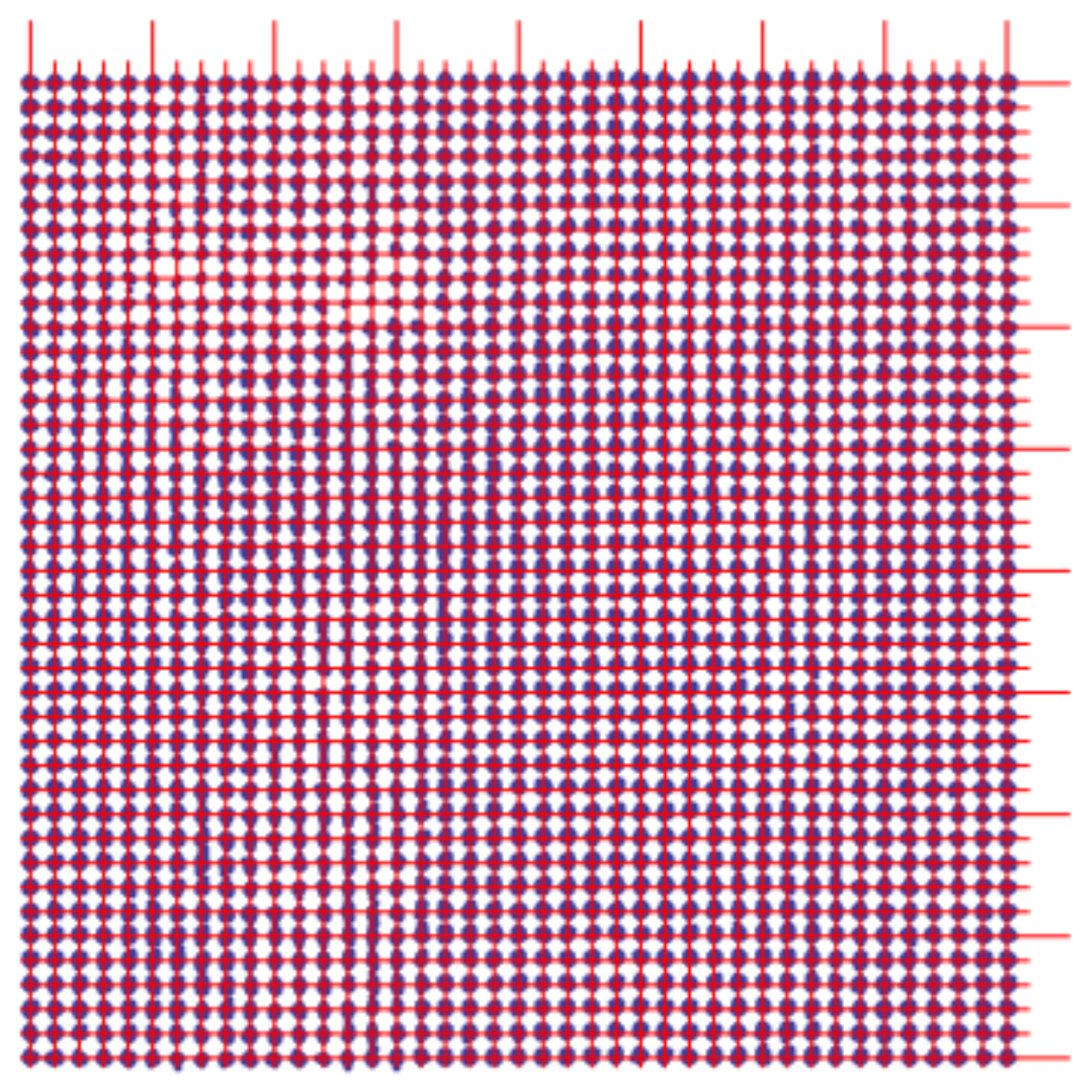}
\caption{Peaks in the density profile associated with a two-dimensional projection of the centers of mass of the cubes at packing fraction $\eta  = 0.52$, as measured in an EDMD simulation initialized with $N=$40$^3$ particles on a $N_L=$40$^3$ 
simple cubic lattice, i.e. no vacancies.  The number of lattice sites in the system spontaneously increased to $N_L=41^3$ lattice sites, corresponding to a net vacancy concentration of approximately 8\%. The lines were added to highlight the 41 evenly spaced layers in both (x and y) directions.   
\label{fig:density} }
\end{center}
\end{figure}

Looking in detail at the EDMD simulations for $\eta$ between 0.52 and 0.56, i.e., in the region where 
Ref. \cite{escobedopolyhedral} reported the cubatic phase, 
we noticed that in many simulations the crystal lattice spontaneously transformed in one of two distinct ways (see the supporting information). 
In most cases, the simple cubic crystal did not maintain its original 
orientation in the box; instead it rotated, introducing defects and frustrations to the crystal lattice.  In others, the 
system spontaneously added extra layers, i.e. extra lattice sites.
As long as the crystal lattice remains aligned with the simulation box, the number of lattice sites can easily be measured from the number of 
peaks in the three-dimensional density profile of the cubes.  Figure \ref{fig:density} shows a two-dimensional 
projection of such a density profile from simulations with $N=40^3$ particles. From this plot we can determine that the system has $N_L=41^3$ lattice sites. 
Thus, the system spontaneously incorporated a large number of  excess lattice sites into the crystal. 
The resulting crystal has a net vacancy concentration, of approximately 
$\alpha=(N_L-N)/N_L=8$\%.
It should be noted that since the volume and the number of particles in the system are fixed,
it is generally not possible to reach an equilibrium concentration of defects in the system. However, 
the formation of extra layers and lattice distortions both significantly increase the number of lattice sites 
in the crystal, and suggest that the thermodynamically stable phase in this regime might be a vacancy-rich crystal structure. 
We note here that the systems sizes examined by Ref. \cite{escobedopolyhedral} would not have allowed for extra layers to form.  
Hence, we believe that the rotated, defective crystals we see are what the authors of Ref. \cite{escobedopolyhedral} identified as cubatic.

\begin{figure}
\begin{center}
\includegraphics[width= 0.45\textwidth]{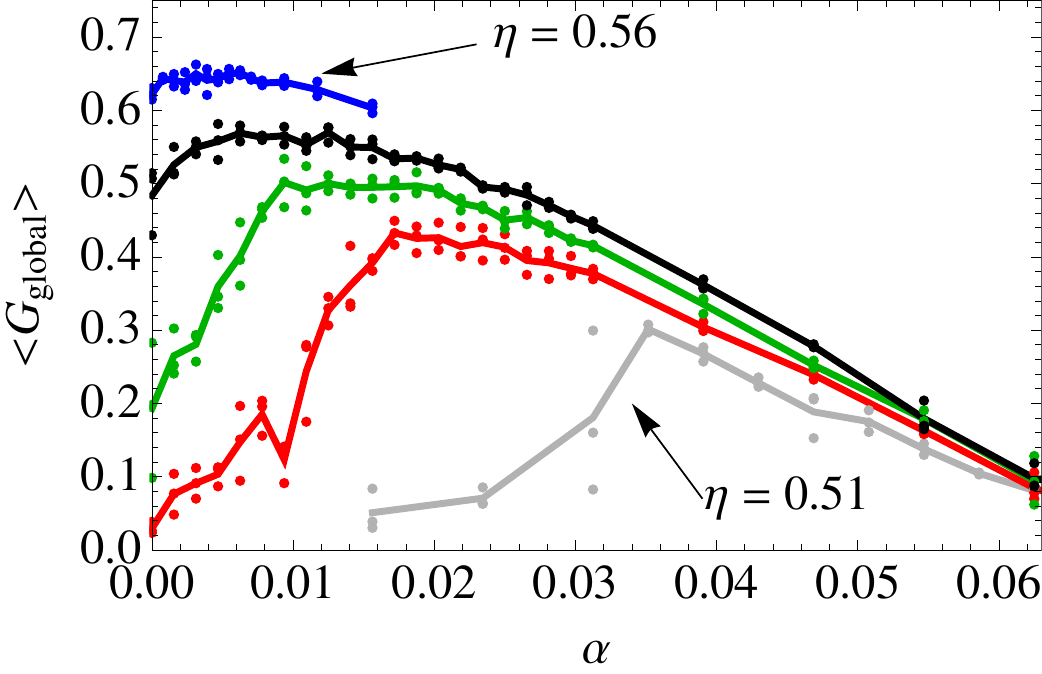}
\caption{Global positional order $\left< G_{\mathrm{global}} \right>$ as a function of the net vacancy concentration $\alpha$ for packing fractions $\eta=0.51$, 0.52, 0.53, 0.54, and 0.56 in a system with $N_L=64000$ lattice sites.  The points indicate
measurements of  $\left< G_{\mathrm{global}} \right>$ along the $x$, $y$, and $z$ axes separately, while the 
lines are averaged over all three directions.  
 \label{fig:order}}
\end{center}
\end{figure}

To examine the effect of vacancies on the crystal structure, we performed additional EDMD simulations on systems 
with various net vacancy concentrations $\alpha = (N_L-N)/N_L$ for a number of packing fractions $\eta=N \sigma^3/V$. 
The average global positional order in the system was measured using the 
positional order parameter averaged over all particles:
\begin{equation}
\left< G_{global} \right> = \left| \left( 1/N \right)\sum_j\exp(i \mathbf{K}\cdot\mathbf{r}_j) \right|,
\end{equation}
where $\mathbf{K}$ is a reciprocal lattice vector of the crystal under consideration
and $\mathbf{r}_j$ is the position of particle $j$.  In all our plots we have chosen $\mathbf{K}$ to correspond
to a single lattice vector, i.e $\mathbf{K_\mathbf{e}}=\frac{2\pi}{a} \hat{\mathbf{e}}$ with $\hat{\mathbf{e}} = \hat{\mathbf{x}},\hat{\mathbf{y}},\hat{\mathbf{z}}$
and $a$ the lattice spacing. 
To set the net vacancy concentration we placed $N=(1-\alpha) N_L$ particles randomly on a simple cubic lattice with $N_L=40^3=64000$ lattice sites, 
and then rescaled the volume (and thus the lattice spacing) of the box to the desired packing fraction $\eta$. Hence, the resulting system
has a lattice spacing that depends on the chosen packing fraction $\eta$ and net vacancy concentration $\alpha$. 
These simulations show that there is a maximum in the global positional order as a function of net vacancy 
concentration $\alpha$ for varying packing fractions $\eta$ (Fig. \ref{fig:order}).  
An increase in order due to an increase in number of vacancies is unexpected as typically the presence of 
defects (such as vacancies) decreases the order. This observation suggests that adding defects reduces 
frustration in the crystal, potentially stabilizing a defect-rich crystal.


\subsection{Defect realization}
\begin{figure}
 \includegraphics[width= 0.45\textwidth]{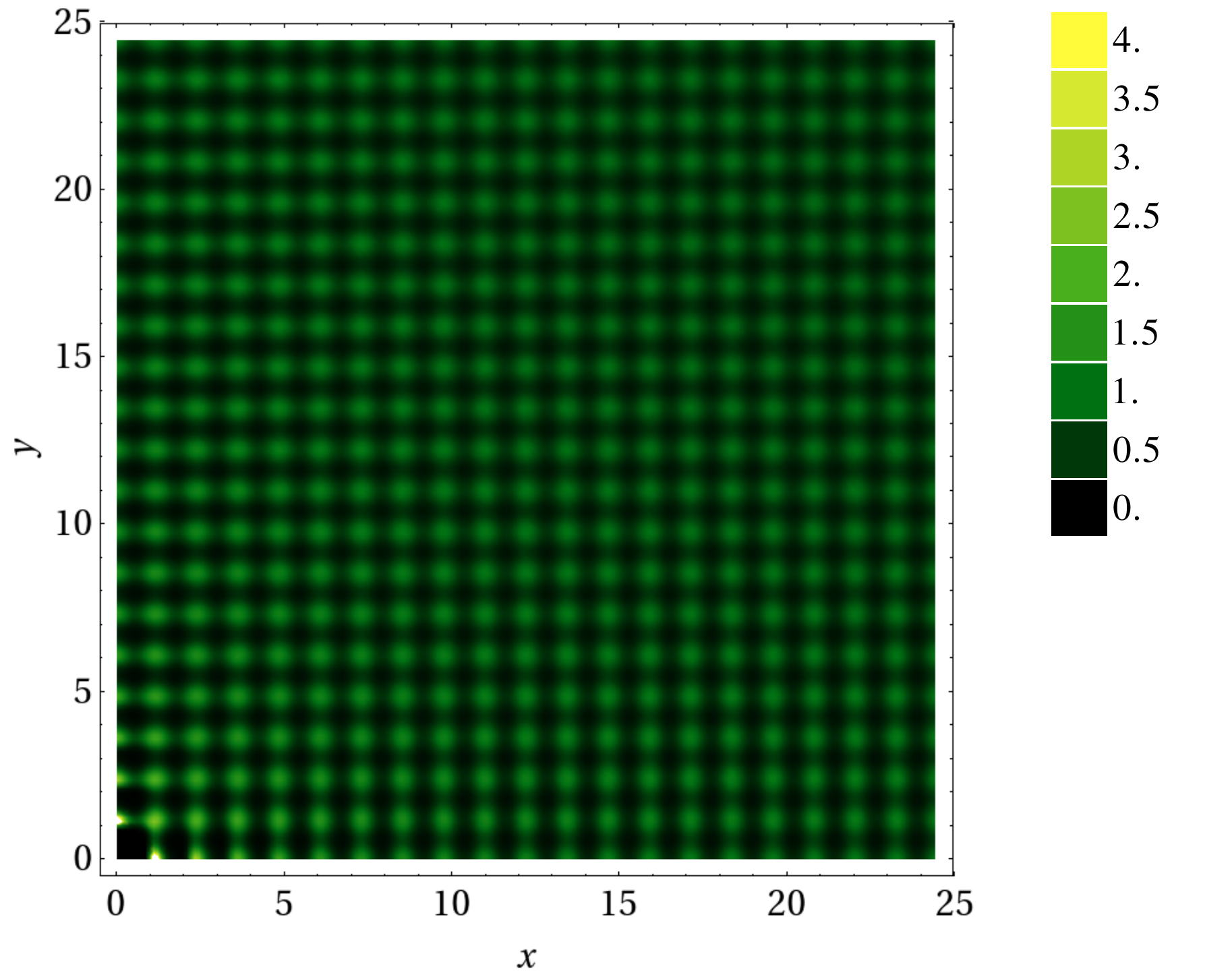}
 \caption{Three-dimensional pair correlation function $g(x,y,z)$ with $z=0$, measured in an EDMD simulation of a system of $N=64000$ particles with packing fraction $\eta = 0.51$ and vacancy concentration $\alpha = 0.055$. The $x$, $y$, and $z$ directions are taken along the three axes of the simulation box. The data is averaged over 50 snapshots, and over the four quadrants of the $xy$-plane.
 \label{gofxy}}
 \end{figure}

\begin{figure}[t]
\begin{center}
\includegraphics[width= 0.3\textwidth]{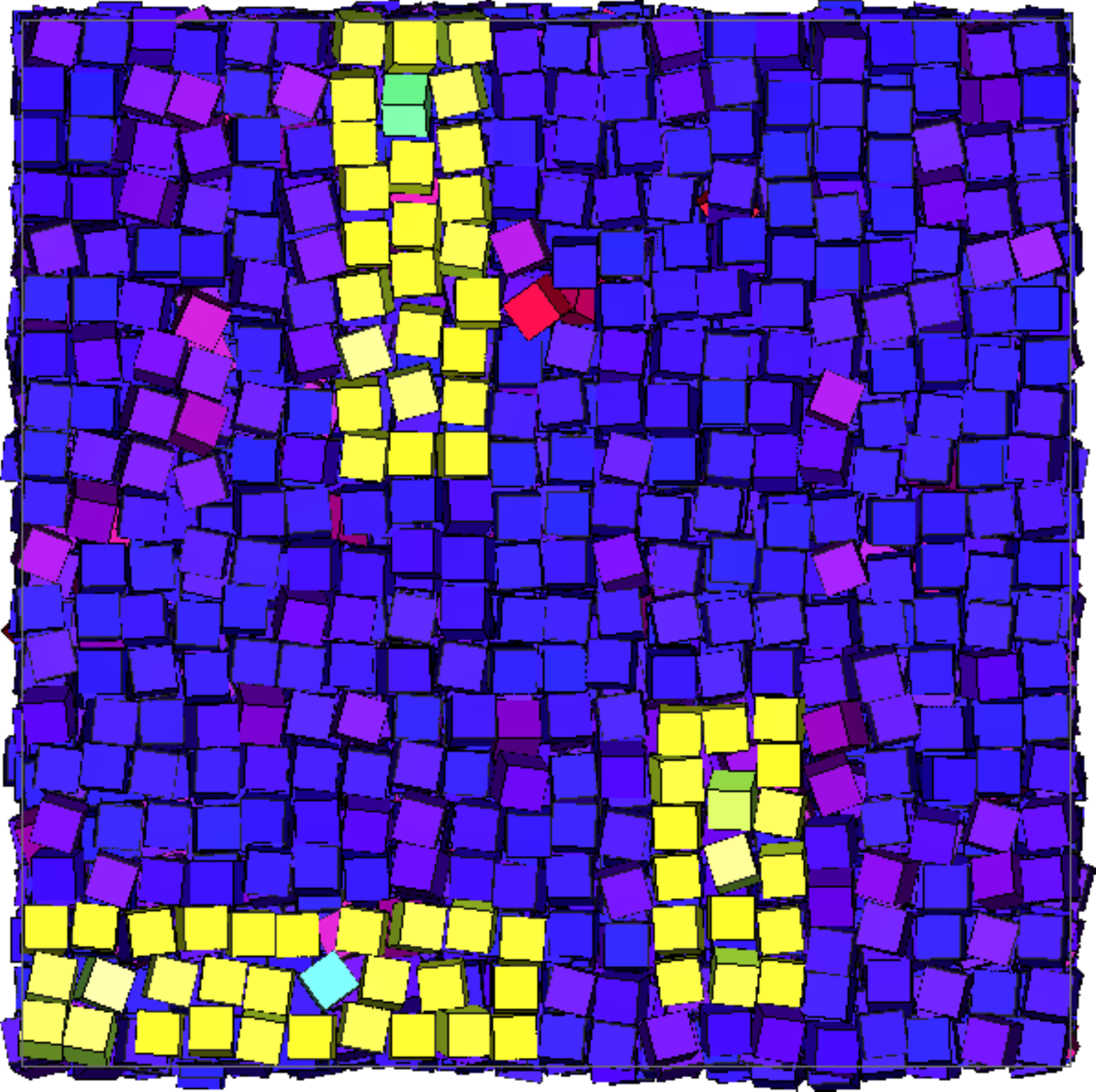}
\end{center}
\caption{A typical configuration where three delocalized defects, and the particles directly surrounding the defects, have been highlighted (yellow) in a system with $N_L = 8000$ lattice sites at packing fraction $\eta=0.56$ and defect concentration $\alpha = 0.016$ from an EDMD simulation. 
The color of the other particles indicates the orientation of each particle with respect to the crystal lattice, with colors closer to blue indicating more closely aligned particles. 
In the defect furthest to the right, the highlighted area shows three cubes sharing four lattice sites.  The uppermost defect has six cubes sharing seven lattice sites, and the bottom most defect has seven cubes sharing eight lattice sites. \label{fig:defectsnapshot}}
\end{figure}

The `net vacancy concentration' $\alpha$, defined above, is simply the excess of lattice sites $N_L$
compared to the number of particles $N$, divided by $N_L$, i.e. the fraction of lattice sites that does not have a particle associated with it.
In a typical system, for instance hard spheres, a vacancy is localized to a single lattice site
and one can determine the number of vacancies by counting the number of 
empty lattice sites. In a hard-sphere crystal, a particle next to an empty lattice site is kept in place by its other neighbors.
This is not the case for hard cubes and, as a result, the way vacancies manifest in this system is very atypical. 
In hard cubes, entirely empty lattice sites are rarely seen even in the vacancy-rich crystals
near coexistence. Instead, a defect manifests itself as a finite-length chain of particles along one of the 
three principal axes in the crystal, in which the particles have a slightly larger inter-particle spacing than the average 
as shown in Fig. \ref{fig:defectsnapshot}.  Hence, if a vacancy extends over 4 lattice sites, as
is the case for one of the vacancies highlighted in Fig. \ref{fig:defectsnapshot}, then the vacancy is realized by 
3 particles sharing 4 lattice sites in a one-dimensional chain. Additionally, while a two dimensional 
layer in a typical snapshot, such as in Fig. \ref{fig:defectsnapshot} shows regions of disorder,
it should be noted that even at high vacancy concentrations, 
the crystal still shows a well-defined lattice spacing on average, which can be easily determined from the 
position of the peaks in the scattering function $S(\mathbf{k})$ or the three-dimensional pair correlation function $g(\mathbf{r})$ (Fig. \ref{gofxy}).

It should be noted that the {\it net} vacancy concentration includes both vacancies as well as interstitials in the sense
that each interstitial cancels a vacancy.  However, the number of vacancies is  higher than the number of interstitials
resulting in the large positive net vacancy concentrations we find in this system.  Similar to a vacancy, interstitials are also not localized in this
system and occur by $n$ particles sharing $n-1$ lattice sites.


\subsection{Phase diagram of hard cubes}
So far we have established a relationship between the order in the system and the net vacancy concentration.  However, 
there is no way to determine the equilibrium net vacancy concentration  and the phase boundaries from the EDMD simulations. 
A proper determination of the equilibrium concentration of vacancies as well as the phase diagram requires 
free-energy calculations. The free energy per particle ($f=\beta F/N$) of the solid with vacancies is given by:
\begin{equation}
f^{\mathrm{vac}}_{\mathrm{ein}}(\lambda) =  f_{\mathrm{ein}}(\lambda) +  f_{\mathrm{rot}}(\lambda) + f_{\mathrm{comb}},
\end{equation}
where the first term is the translational free energy of a normal Einstein crystal \cite{bookfrenkel}, the second term is the 
rotational free energy of the crystal \cite{noya_2007},  and the third term is the 
combinatorial entropy associated with placing $N$ particles on $N_L$ lattice sites:
\begin{equation}
f_{\mathrm{comb}}=- ({1}/{N}) \log [{N_L!}/({N! (N_L - N)!})].
\end{equation}
Detailed descriptions on how these terms were calculated using Monte Carlo simulations are given in the Methods section.

\begin{figure}
\begin{center}

\includegraphics[height=0.755\columnwidth]{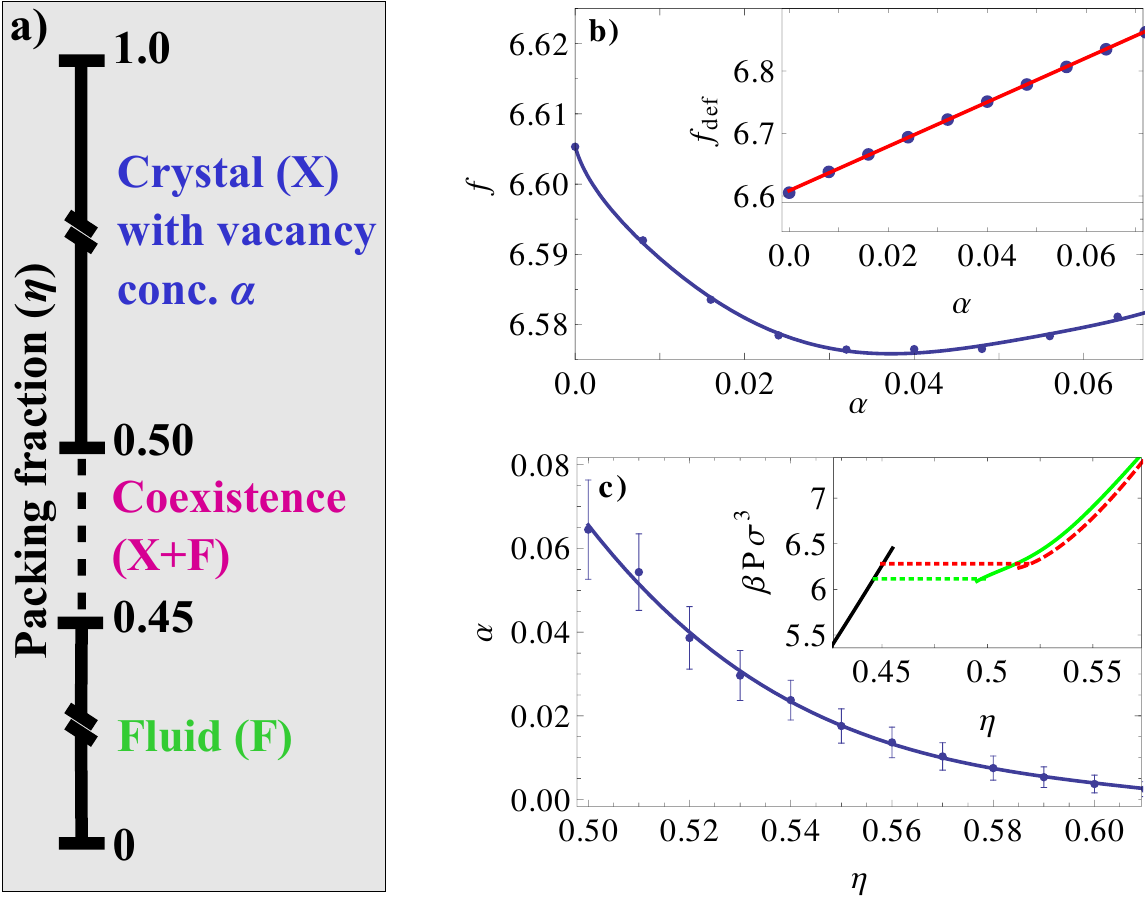}

\caption{{\bf (a)} The equilibrium phase diagram of hard cubes as a function of packing fraction $\eta$. For $\eta<0.45$ we predict a stable fluid
phase, for $\eta>0.50$ we find a stable crystal phase with vacancies (see, e.g. Fig. 2), and in between these two packing fractions we find coexistence between the 
crystal and fluid.  {\bf (b)} Free energy per particle as a function of net vacancy concentration $\alpha$ at packing fraction $\eta = 0.52$. The points correspond to measurements while the solid line is a guide to the eye. The estimated error in the free energies, based on independent runs, is $\simeq 0.004 k_B T$. Inset: Free energy per particle without taking into account the combinatorial free energy: $f_\mathrm{def} = f - f_\mathrm{comb}$. The solid line is a linear fit.
{\bf (c)} The net vacancy concentration $\alpha$ as a function of packing fraction $\eta$ in the crystal phase. The error bars are based on the width of the free energy minimum.
 Inset: Equations of state for the fluid phase (black), the stable vacancy-rich crystal (green), and the crystal without vacancies, i.e. $\alpha = 0$ (red, dashed). Note that the phase transition (dotted lines) shifts to lower densities when vacancies are taken into account. The phase transition for the vacancy-free system essentially coincides with the one in Ref. \cite{escobedopolyhedral}.
 \label{fig:defects}}
\end{center}
\end{figure}

We  determined the free energy as a function of net vacancy concentration $\alpha$, at a fixed packing fraction $\eta=0.52$.  As shown in the inset in Fig. \ref{fig:defects}b, the minimum in the free energy occurs for a high concentration of vacancies. 
Specifically, we find that at this density the number of particles is  4\% lower than the number of lattice sites.  
While calculating the free energy as a function of $\alpha$, we also 
observed that the free energy, excluding the combinatorial contribution, was  almost linear. 
This is shown in the inset of Fig. \ref{fig:defects}b.
While we are uncertain to the origin of this linearity, it seems to suggest that the vacancies 
are only weakly interacting. 

For each value of $\alpha$, the free energy as a function of density was obtained by combining the reference free energies shown in Fig. \ref{fig:defects}b with a separate equation of state measured for that value of $\alpha$.
By minimizing the resulting free energy with respect to $\alpha$, we find that the number of excess lattice sites decreases as a function of 
the density, as expected (Fig. \ref{fig:defects}c).


Using a common tangent construction (see the supporting information) in 
combination with the determined free energies, we mapped out the phase diagram which is shown in Fig. \ref{fig:defects}a.
We find coexistence between a fluid phase with coexisting density $\eta_f=N\sigma^3/V=0.45$ and a vacancy-rich simple cubic
crystal structure with coexisting density $\eta_m=0.50$ and net vacancy concentration $\alpha = (N_L - N)/N_L \simeq 0.064$. The
pressure and chemical potential at coexistence are $\beta p \sigma^3 = 6.16$ and $\beta \mu = 18.4$, respectively.
The inset of Fig. \ref{fig:defects}c shows the equations of state and phase transitions both including and excluding the effects of vacancies in the crystal phase.
The presence of vacancies in the crystal significantly lowers the melting density compared to the one reported by 
Ref. \cite{escobedopolyhedral} where they found that a defect-free crystal melted at
$\eta_m=0.52$, while the freezing packing fraction is approximately the same. We note here 
that we also find a melting number density of $\eta_m=0.52$ if we exclude vacancies.
Hence, we find that vacancies increase the range of stability of the simple cubic crystal.


\subsection{Diffusion}
\begin{figure}
\begin{center}
\includegraphics[width= 0.45\textwidth]{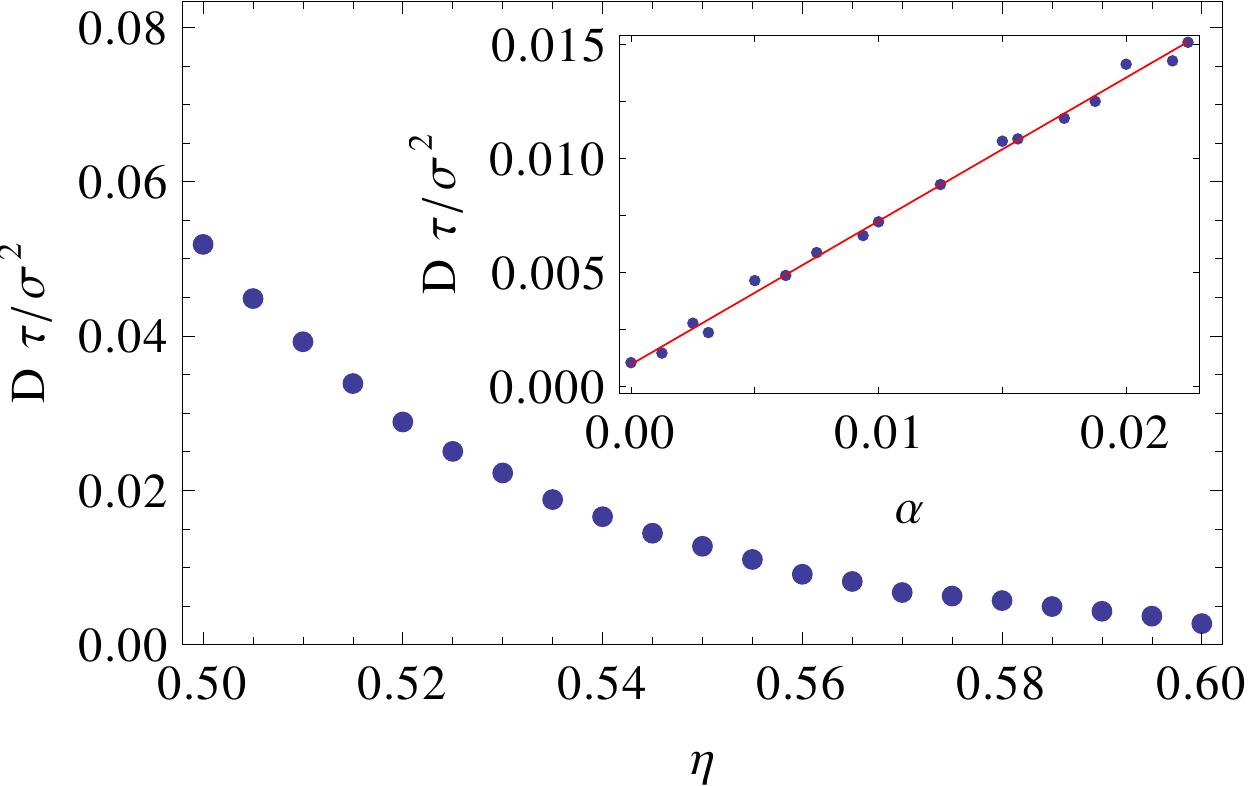}
\caption{Dimensionless diffusion coefficient $D \tau / \sigma^2$ in the solid phase as a function of the packing fraction $\eta$. Here, $\tau = \sqrt{\beta m \sigma^2}$ is the unit of time in the EDMD simulations and $m$ is the mass of a single cube. For each density, the crystal has the equilibrium net vacancy concentration as determined from the free energy calculations. The inset shows the diffusion coefficient at a fixed packing fraction $\eta = 0.56$ and varying net vacancy concentration.
\label{fig:diffusion} }
\end{center}
\end{figure}

As was already shown in Ref. \cite{escobedopolyhedral}, the ordered phase has appreciable diffusion in the intermediate density regime. 
This can be understood in terms of the delocalized defects, which diffuse through the crystal and allow particles to diffuse in the opposite direction.
To investigate the effect of vacancies on the diffusion coefficient in the solid, we measured the long-time self-diffusion constant of cubes in the crystal phase using EDMD simulations. Figure \ref{fig:diffusion} shows the diffusion constant as a function of density, where the net vacancy concentration at each density was chosen to correspond to the equilibrium net vacancy concentration shown in Fig. \ref{fig:defects}c. Near coexistence, the diffusion coefficient increases significantly, up to a maximum of $D \tau /\sigma^2 = 0.05$, where $\tau = \sqrt{\beta m \sigma^2}$ is the unit of time in the EDMD simulations and $m$ is the mass of a single cube. For comparison, the diffusion constant in the fluid at coexistence is $D \tau / \sigma^2  = 0.15$, three times higher than in the coexisting solid.

At fixed density, the diffusion constant increases approximately linearly with the number of vacancies, with very little diffusion remaining at $\alpha = 0$. An example of this is shown in the inset of Fig. \ref{fig:diffusion}, for packing fraction $\eta = 0.56$. Note that even for vanishing net vacancy concentration, diffusion is still possible via the spontaneous formation of delocalized interstitial-vacancy pairs. However, at the equilibrium net vacancy concentration ($\alpha = 0.013$), the diffusion coefficient is eight times as high as in the vacancy-free crystal, indicating that vacancies play a major role in the dynamics of the particles in the solid.

\section{Discussion and Conclusions}
In this paper, we have examined the effects of vacancies on the phase diagram of hard cubes
using both event driven molecular dynamics simulations as well as Monte Carlo simulations. 
From the molecular dynamics simulations it is clear that vacancies play an important role in 
the equilibrium phase behavior of hard cubes.  Free-energy calculations 
show conclusively a first-order phase transition between
a fluid phase and a vacancy-rich simple cubic crystal phase with up to 6\% vacancies.
Up to the system sizes we have studied (40$^3$ particles), we 
find long-range positional order for systems with an equilibrium concentration of vacancies (see Fig. \ref{gofxy}).  
Thus, we find that the stable phase is a simple cubic crystal for all densities, albeit one with significant diffusion due to the high defect concentration.

The number of vacancies in this system is orders of magnitude larger than typically seen in colloidal systems.
The stability of this vacancy-rich phase can most likely be attributed to the delocalization of defects in the crystal (Fig. \ref{fig:defectsnapshot}):
clearly, one vacancy provides additional free volume for multiple nearby particles, decreasing the entropic cost of creating
a defect. In most other colloidal crystals, such as hard-sphere face-centered cubic crystals, particles near a vacancy are still confined to their lattice site by
their remaining neighbors. As a result, the local entropy gain from a defect is much lower.
The only other similar result we are aware of is for parallel hard cubes;
a somewhat artificial system which shows a very peculiar second order freezing transition from a fluid to 
a simple cubic crystal. 
\cite{Cuesta,Groh_Mulder_2001,Matthieu}


The existence of a vacancy-stabilized simple cubic phase in hard cubes leads 
to the question of whether vacancy-stabilized crystal structures are present 
in other anisotropic, entropy-driven systems.
We would expect vacancies to be relevant for other (likely hard) systems with crystal structures where vacancies can delocalize. Note that delocalization also requires the absence of strong interactions that constrain particles to their lattice sites. For example, we would not expect high vacancy concentrations to occur in the simple cubic structure studied by Rechtsman {\it et al.},\cite{interactingsc} which resulted from isotropic interactions.
Recently, the phase behavior of a large number of polyhedral shapes has been studied using Monte Carlo simulations. \cite{escobedopolyhedral, glotzer1, glotzer2} Since vacancies are easily overlooked in the case of spontaneously formed crystals, and unlikely to form in simulations starting from a fully filled lattice, it is possible that high equilibrium vacancy concentrations occur in many of these systems. 
Specifically, for crystal structures where some or all of the neighboring particles can freely move into an empty lattice site, the possibility of crystal vacancies should likely be taken into account. Examples include the crystal structures predicted for hexagonal and triangular prisms in Ref. \cite{escobedopolyhedral}, or the 2d 
(rounded) hard squares studied in Refs. \cite{hardsquares, escobedosquares}.




\section{Methods}
\subsection{Event driven molecular dynamics simulations}  Please refer to the supporting text for a full description.

\subsection{Free energy of the liquid}
Thermodynamic integration allows one to calculate the free energy for all densities assuming that both the equation of state and the free energy at a reference density are known. 
When the free energy of a reference density $F(\rho_0)$ is known, the free energy as a
function of number density $F(\rho)$ can be determined using the equation of state. In particular, the free energy is given by
\begin{equation}
\frac{\beta F\left( \rho \right)}{N} = \frac{\beta F\left( \rho_0 \right)}{N}
+ \beta \int_{\rho_0}^{\rho} \frac{P\left(\rho^{\prime}\right)}{\left(\rho^{\prime}\right)^2}d\rho^{\prime} \label{eq:thermoint}
\end{equation}
where $\rho$ is the density and  $\beta = 1/k_B T$ with $k_B$ Boltzmann's constant and $T$ the temperature.
To measure the free energy of the fluid at a reference density, we used Widom insertion test particle method\cite{bookfrenkel}. 
The free energy of the fluid at density $\rho_0$ is then given by
\begin{equation}
 \frac{\beta F_\mathrm{f}(\rho_0)}{N} = \beta \mu(\rho_0) - \frac{\beta P(\rho_0)}{\rho_0}
\end{equation}

\subsection{Solid free energies with and without vacancies}
To calculate the Helmholtz free energy as a function of the density for the solid phase
we use thermodynamic integration \cite{bookfrenkel} in MC simulations of systems with $N_L = 20^3 = 8000$ or $N_L = 25\times 20\times 18=9000$ lattice sites. We checked during our simulations that the number of lattice sites did not change spontaneously. For systems with the same density and net vacancy concentration, the differences in free energy between these two lattice sizes were within the error of our measurements. However, while equivalent free energy calculations for a smaller system ($N_L = 1000$ lattice sites) yielded qualitatively similar results, finite-size effects were noticeable when compared to the larger systems.

For the reference free energy of a crystal without vacancies, 
we use a variation on the method introduced by Frenkel and Ladd \cite{bookfrenkel}, where 
particles are tied to their respective lattice sites with springs, transforming the crystal into 
a non-interacting Einstein crystal for a sufficiently high spring constant $\lambda$. In this case, 
we also add an aligning potential to handle the orientational degrees of freedom of 
the particles \cite{noya_2007}. Using the same coupling constant $\lambda$ that attaches the 
particles to their lattice sites, the aligning potential is given by 
\begin{equation}
\beta U_{\mathrm{rot}}(\lambda) = 
\lambda \sum_{i=1}^{N}  \min_{j \ne k} 
\left\{ 2-\left(\mathbf{u}_{i,j} \cdot \hat{\mathbf{x}} \right)^2-\left(\mathbf{u}_{i,k} \cdot \hat{\mathbf{y}} \right)^2 \right\} 
\end{equation}
where $\hat{\mathbf{x}}(\hat{\mathbf{y}})$ is a unit vector along the $x(y)$ axis, 
and $\mathbf{u}_{i,j}$, with $j=1,2,3$, are three mutually perpendicular 
face normals associated with particle $i$. Also, $\beta=1/k_B T$ is the inverse thermal energy, where $k_B$ is Boltzmann's constant and $T$ the temperature. 
The parameter $\lambda$ controls the strength of the external potentials; 
hence for $\lambda=0$ the system reduces to pure hard cubes, and 
for  $\lambda=\lambda_m$ with $\lambda_m$ sufficiently large, the particles in the crystal are non-interacting.  

To calculate the free energy of a system with vacancies, instead of fixing the particles to a specific lattice site, 
we attach the particles to their nearest lattice site \cite{multipleoccupancy} using 
\begin{equation}
U_{\mathrm{ext}}(\lambda) = 
  \lambda \sum_{i=1}^{N}  \left(\frac{1}{\sigma^2} \left|\mathbf{r}_i - \mathbf{r}^0(\mathbf{r}_i) \right|^2  \right) + U_{\mathrm{rot}}(\lambda)
\end{equation}
where  $\mathbf{r}^0(\mathbf{r}_i)$ is the position of the lattice site nearest to $\mathbf{r}_i$. 
In this case, the dimensionless free energy per particle ($f=\beta F/N$) of the noninteracting system is
$f^{\mathrm{vac}}_{\mathrm{ein}}(\lambda) =  f_{\mathrm{ein}}(\lambda) +  f_{\mathrm{rot}}(\lambda) + f_{\mathrm{comb}}$
where the first term is the translational free energy of a normal Einstein crystal \cite{bookfrenkel}, the second term is the 
rotational free energy of the crystal \cite{noya_2007},  and the third term is the 
combinatorial entropy associated with placing $N$ particles on $N_L$ lattice sites:
$f_{\mathrm{comb}}=- ({1}/{N}) \log [{N_L!}/({N! (N_L - N)!})]$.
The full free energy of the crystal of hard cubes with vacancies is then given by:
\begin{equation}
f = f^{\mathrm{vac}}_{\mathrm{ein}}(\lambda_m) - \frac{\beta}{N} \int_0^{\lambda_m} \left< \frac{\partial U_\mathrm{ext}(\lambda')}{\partial \lambda'} \right>_{\lambda'} \mathrm{d}\lambda'.
\end{equation}
In contrast to the free energy calculations for systems without vacancies \cite{bookfrenkel}, 
the center of mass of the system  is not fixed in these simulations. 
To equilibrate the position of the center of mass, we introduce MC moves that collectively translate every 
particle in the system \cite{multipleoccupancy}. Additionally, moves that translate a single particle by exactly one lattice 
vector are introduced in order to improve sampling of different distributions of vacancies over the crystal. For a 
system with full lattice site occupancy ($N = N_L$) and thus no vacancies, we obtain good agreement between the two methods.

\subsection{Diffusion constants}
To measure the long-time self-diffusion constant in the crystalline phase shown in Fig. \ref{fig:diffusion}, we performed EDMD simulations in systems of $N_L = 8000$ lattice sites, for a range of densities. The vacancy concentration was chosen to correspond to the equilibrium vacancy concentration shown in Fig. \ref{fig:defects}c. The diffusion constant was calculated from the slope of the mean squared displacement as a function of time. An example of a plot showing the mean squared displacement is shown in the supporting text.



\begin{acknowledgments}
The authors thank M. Miller and D. Frenkel for 
many useful discussions.  LF acknowledges support from the EPSRC, U.K. for funding (Programme Grant EP/I001352/1), FS and MD acknowledge support of a NWO-VICI grant. MM acknowledges support of the SFB-TR6 program (project D3).
\end{acknowledgments}


\begin{thebibliography}{10}

\bibitem{defects}
S. Pronk and D. Frenkel, 
Large effect of polydispersity on defect concentrations in colloidal crystals,
\textit{J. Chem. Phys.}, 120:6764  (2004).

\bibitem{microcubesynthesis1}
T. Sugimoto and K. Sakata,
Preparation of monodisperse pseudocubic [alpha]-Fe2O3 particles from condensed ferric hydroxide gel,
\textit{J. Coll. Int. Sci.},  152:587   (1992).

\bibitem{nanocubesynthesis1}
C.~J. Murphy, T. K. Sau, A. M. Gole, C. J. Orendorff, J. Gao, L. Gou, S. E. Hunyadi, and T. Li,
Anisotropic Metal Nanoparticles:  Synthesis, Assembly, and Optical Applications,
\textit{J. Phys. Chem. B}, 109:13857  (2005).

\bibitem{nanocubesynthesis2}
Z. Pu, M. Cao, J. Yang, K. Huang and C. Hu,
Controlled synthesis and growth mechanism of hematite nanorhombohedra, nanorods and nanocubes,
\textit{Nanotechnology}, 17:799  (2006).

\bibitem{nanocubesynthesis3}
C. Su, H. Wang, and X. Liu, 
Controllable fabrication and growth mechanism of hematite cubes,
\textit{Crystal Research and Technology}, 46:209 (2011).

\bibitem{rossi2010cubic}
L. Rossi, S. Sacanna, W. T. M.  Irvine, P. M.  Chaikin, D. J. Pine, and A. P. Philipse,
Cubic crystals from cubic colloids,
\textit{Soft Matter} 7:4139  (2011).

\bibitem{nanocubes2011}
Y. Zhang, F. Lu, D. van der Lelie and O. Gang,
Continuous Phase Transformation in Nanocube Assemblies,
\textit{Phys. Rev. Lett.} 107:135701  (2011).

\bibitem{edmdanisotropic}
L. H. de la Pe\~{n}a, R. van Zon, J. Schofield and S. B. Opps,
Discontinuous molecular dynamics for semiflexible and rigid bodies,
\textit{J. Chem. Phys.}  126:074105  (2007).

\bibitem{separatingaxis}
S. Gottschalk, M.~C. Lin, and D. Manocha,
OBBTree: a hierarchical structure for rapid interference detection,
\textit{Proc. ACM SIGGRAPH}, 171  (1996).

\bibitem{escobedopolyhedral}
U. Agarwal and F. Escobedo, 
Mesophase behaviour of polyhedral particles,
\textit{Nature Materials} 10:230  (2011).

\bibitem{escobedoparallelepipeds}
B.~S. John, C. Juhlin, and F.~A. Escobedo,
Phase behavior of colloidal hard perfect tetragonal parallelepipeds,
\textit{J. Chem. Phys.}, 128:044909 (2008).

\bibitem{jagla_1998}
E.~A. Jagla,
Melting of hard cubes,
\textit{Phys. Rev. E}, 58:4701  (1998).

\bibitem{bookfrenkel}
D. Frenkel and B. Smit, 
Understanding Molecular Simulations: From Algorithms to Applications (Academic Press, London, UK, 2002).

\bibitem{noya_2007}
E. G. Noya,  C. Vega,  J. P. K. Doye and A. A. Louis,
Phase diagram of model anisotropic particles with octahedral symmetry, 
\textit{J. Chem. Phys.} 127:054501  (2007).

\bibitem{multipleoccupancy}
B. M. Mladek, P. Charbonneau, C. N. Likos, D. Frenkel and G. Kahl,
Multiple occupancy crystals formed by purely repulsive soft particles,
\textit{J. Phys: Cond. Mat.} 20:494245  (2008).

\bibitem{Cuesta}
Yuri Mart{\'i}nez-Raton and Jos{\'e} A. Cuesta,
Fundamental measure theory for mixtures of parallel hard cubes. II. Phase behavior of the one-component fluid and of the binary mixture,
\textit{J. Chem. Phys.} 111:317 (1999).

\bibitem{Groh_Mulder_2001}
B. Groh and B. Mulder,
A closer look at crystallization of parallel hard cubes,
\textit{ J. Chem. Phys.} 114:3653  (2001).


\bibitem{Matthieu}
M. Marechal, U. Zimmermann and H. L{\"o}wen, 
Freezing of parallel hard cubes with rounded edges,
\textit{J. Chem Phys} 136.144506 (2012).

\bibitem{interactingsc}
M. C. Rechtsman, F. H. Stillinger and S. Torquato,
Self-assembly of the simple cubic lattice with an isotropic potential,
\textit{Phys. Rev. E}, 74:021404 (2006).

\bibitem{glotzer1}
P. F. Damasceno, M. Engel and S. C. Glotzer,
Structural diversity and the role of particle shape and dense fluid behavior in assemblies of hard polyhedra,
\textit{arXiv} 1202.2177 (2012).

\bibitem{glotzer2}
P. F. Damasceno, M. Engel and S. C. Glotzer,
Crystalline assemblies and densest packing of a family of truncated tetrahedra and the role of directional entropic forces,
\textit{ACS Nano} 6:609 (2012).


\bibitem{hardsquares}
K.~W. Wojciechowski and D. Frenkel,
Tetratic phase in the planar hard square system?,
\textit{Comput. Methods Sci. Technol.} 10:235  (2004).


\bibitem{escobedosquares}
C. Avenda{\~n}o and F. A. Escobedo,
Phase behavior of rounded hard-squares,
\textit{Soft Matter} 8:4675 (2012)



\end{thebibliography}




\end{document}